\documentclass[conference,14pt]{IEEEtran}
\input epsf
\usepackage{times,mathptm,amsmath}
\usepackage{graphicx}
\usepackage{epsfig}
\usepackage{enumerate}

{\catcode`\:=\active \catcode`\/=\active \catcode`\.=\active
\catcode`\-=\active \catcode`\@=\active
\gdef\url{\tt\catcode`\/=\active \catcode`\:=\active
\catcode`\.=\active \catcode`\-=\active \catcode`\@=\active
\def:{\char`\:\discretionary{}{}{}}%
\def/{\char`\/\discretionary{}{}{}}%
\def.{\char`\.\discretionary{}{}{}}%
\def-{\char`\-\discretionary{}{}{}}%
\def@{\char`\@\discretionary{}{}{}}}}

\title{Practical Resource Allocation Algorithms for QoS in OFDMA-based Wireless Systems}

\author{
\authorblockN{Tolga Girici\authorrefmark{2}, Chenxi Zhu\authorrefmark{1},Jonathan R. Agre\authorrefmark{1}, Anthony Ephremides\authorrefmark{2}}
\vspace{0.02in}
\authorblockA{\authorrefmark{2}
Institute for Systems Research\\
A.V.William Building\\
University of Maryland\\
Email:\{tgirici,etony\}@eng.umd.edu}
\authorblockA{\authorrefmark{1}
Fujitsu Labs of America\\
8400 Baltimore Ave., Suite 302\\
College Park, Maryland 20740\\
Email:\{czhu,jagre\}@fla.fujitsu.com}}

\date{}

\begin{document} \maketitle

\begin{abstract}
In this work we propose an efficient resource allocation algorithm
for OFDMA based wireless systems supporting heterogeneous traffic.
The proposed algorithm provides proportionally fairness to data
users and short term rate guarantees to real-time users. Based on
the QoS requirements, buffer occupancy and channel conditions, we
propose a scheme for rate requirement determination for delay
constrained sessions. Then we formulate and solve the proportional
fair rate allocation problem subject to those rate requirements and
power/bandwidth constraints. Simulations results show that the
proposed algorithm provides significant improvement with respect to
the benchmark algorithm.
\end{abstract} \normalsize

\section{Introduction}

Broadband wireless networks are designed to be able to provide high
rate and heterogenous services to mobile users that have various
quality of service (QoS) requirements.  Two notable examples of
broadband wireless technologies are 3GPP and Mobile WiMax(802.16e).
Transmissions in Long Term Evolution (3GPP) and 802.16-based
wireless technologies are based on OFDM, where several modulation,
coding and power allocation schemes are allowed to give more degrees
of freedom to resource allocation \cite{Gho05}. Fully taking
advantage of this degree of freedom is an important problem and has
been studied previously in \cite{Rhe00, She05, Han05, Son_dis,
Zhu06}. Papers  \cite{Rhe00} and \cite{Han05} address maximizing
total throughput subject to power and subcarrier constraints. Above
works consider maximizing total capacity for data traffic but do not
address fairness for data traffic or QoS for real time traffic. The
authors in \cite{She05}, \cite{Son_dis}, \cite{Zhu06} studied
proportional fair scheduling. However these schemes also do not
guarantee any short or long term transmission rates. The scheduling
rules do not apply sufficiently to different QoS requirements and
heterogeneous traffic.

In OFDMA, a wideband channel is divided into a number of narrow-band
carriers and these carriers are allocated to users. Typically the
carriers that are close in the frequency spectrum have correlated
channel conditions.  In order to make the allocation easier carriers
are grouped into subchannels.  There are various ways of
subchannelization, e.g. contiguous grouping (i.e. Band AMC), where
adjacent carriers are grouped into a single subchannel. By this
method it is safe to assume that each subchannel is subject to
independent and identically distributed fading. This method fully
takes the advantage of OFDMA by frequency selectivity. Another
method is the distributed grouping (i.e. PUSC/FUSC) where a
subchannel is formed by sampling carriers across the whole range of
subcarriers according to a permutation, or randomly, so that each
subchannel has the same average fading with respect to a user. Most
of the previous works has considered the first method in their
models, however it has two main disadvantages for mobile networks.
First, the proposed algorithms become too complex when each
subchannel has different fading. We choose permutational method for
subchannelization. Therefore our question becomes how many
subchannels to allocate instead of which subchannels , which makes
our resource allocation algorithms more practical. Second, for a
mobile channel with fast fading, channel estimation and feedback
becomes more practical using distributed grouping.

Motivated by the above issues we propose a resource allocation
algorithm, that satisfies delay requirements for real time traffic,
while providing proportional fair rate allocation for elastic
traffic. Our algorithm is based on user selection and rate
requirement determination for voice users and solution of a
proportional fair rate allocation problem subject to those rate
requirements and power/bandwidth constraints.

\section{System Model\label{Ch3:system_model}}

We consider a cellular system consisting of a single base station
transmitting to $N$ mobile users. Time is divided into frames of
length $T_f$ and at each time frame base station allocates the total
bandwidth $W$ and total power $P$ among the users.  In the
simulations we keep the users fixed, however we simulate mobility by
fast and slow fading. Fast fading is Rayleigh distributed and slow
fading is log-normal distributed. Total channel gain is the product
of distance attenuation, fast and slow fading. Let $h_i(t)$ be the
channel gain of user i at time t. For an AWGN channel with noise
p.s.d. $N_0$, signal to interference plus noise ratio (SINR) is,
\begin{equation}
SINR_i=\frac{p_i(t)h_i(t)}{N_0w_i(t)},
\end{equation}where $p_i(t)$ and $w_i(t)$ are the power and
bandwidth allocated to user i at time t.

We use the following (modulation,coding,repetition) pairs
[QPSK,1/2,6$\times$ - QPSK,1/2,4$\times$ - QPSK,1/2,2$\times$ -
QPSK,1/2,1$\times$1 - QPSK,3/4,1$\times$ - 16QAM,1/2,1$\times$ -
16QAM,3/4,1$\times$ - 64QAM,2/3, 1$\times$ - 64QAM,3/4,1$\times$]
corresponding to the following SINR levels: [-2.78, -1.0, 2.0, 5, 6,
10.5, 14, 18, 20] dB \cite{80216e}. For instance QPSK,1/2,6$\times$
corresponds to a bandwidth efficiency of 1/6 bps/Hz. In the problem
formulation, we will use the following rate function.
\begin{equation}
r_i(p_i(t),w_i(t))=w_i(t)\log\left(1+\beta\frac{p_i(t)h_i(t)}{N_0w_i(t)}\right),\label{rate_equation}
\end{equation}which is the Shannon capacity expression with an SINR
factor $\beta<1$.  If we choose $\beta=0.25$, this rate function
approximates the above values quite well. After allocating the power
and bandwidth we quantize the SINR to the values above. Bandwidth
also is quantized to multiples of subchannel bandwidth, $W_{sub}$.

The network can support different traffic types such as real time
(VoIP), video streaming, data applications with some rate
requirements (FTP) and best effort traffic. We assume that each user
demands a single type of traffic.   We will consider the following
traffic types: 1) Best Effort (BE): Non real time traffic with no
minimum rate requirements. 2) Video Streaming: Bursty real time
traffic with delay constraint. 3) VoIP: Constant bit rate (CBR)
traffic with delay constraint.

We classify the traffic into two groups as elastic and non-elastic
traffic.  BE traffic is elastic, that is, a BE user can use any
available traffic. Fairness and throughput are the performance
objectives for BE traffic. Proportional fairness provides a good
balances between the two.  Voice traffic is non-elastic; it is a CBR
traffic with strict delay requirements. If a voice user can receive
its short term required rate level, it doesn't need excessive
resources. On the other hand Video streaming traffic is in between
the two types. It has a basic rate requirement with certain delay
constraints, however it is possible to achieve higher quality video
transmission if the user experiences good channel conditions. In
this work we aim to satisfy the basic rate requirement for voice and
video users, while treating excessive rate allocation for video
users similarly as BE users. Typical rates for these traffic types
are listed in Table \ref{parameters}.

\section{User Selection}

Our proposed scheduling algorithm consists of user selection and
rate allocation. After selecting the users, the subchannels and
power is allocated.

\subsection{Modified Largest Delay First - Proportional Fairness}

In single channel systems Largest Weighted Delay First (LWDF) is
shown to be throughput optimal \cite{And01}.  In this scheme at each
frame the user maximizing the following quantity transmits
\begin{equation}a_i D_i^{HOL}(t)r_i(P,W),\label{lwdf}\end{equation}where
$D_i^{HOL}(t)$ is the head of line packet delay and $r_i(P,W)$ is
the channel capacity of user i at frame t (calculated from
(\ref{rate_equation}), where P and W is the fixed transmission power
and channel bandwidth). The parameter $a_i$ is a positive constant.
If QoS is defined as
\begin{equation}P(D_i>D_i^{max})<\delta_i,\end{equation}where
$D_i^{max}$ is the delay constraint and $\delta_i$ is the
probability of exceeding this constraint (typically 0.05), then the
constant $a_i$ can be defined as
$a_i=-\frac{\log(\delta_i)}{D_i^{max}R_i(t)}$ , which is referred to
as M-LWDF-PF \cite{And01}. Here, $R_i(t)$ is the average received
rate. Averaged (filtered) values of long term received rates of
users, which is computed as follows:
\begin{equation}
R_i(t+1)=\alpha_i
R_i(t)+(1-\alpha_i)r_i(p_i(t),w_i(t))\label{average_rate}
\end{equation}

The equation above can be considered as a filter with time constant
$1/(1-\alpha_i)$ for user $i$. The constant $\alpha_i$ should be
chosen such that the average received rate is detected earlier than
the delay constraint in terms of frame durations. We choose 100msec,
400 msec and 1000 msec as the delay constraints of voice, streaming
and BE users.  Converting these values into number of frames of
1msec we get the $\alpha$ values in Table \ref{parameters}.
M-LWDF-PF can be adapted to OFDMA systems as follows. Power is
distributed equally to all subchannels. Starting from the first
subchannel , the subchannel is allocated to the user maximizing
(\ref{lwdf}). Then the received rate $R(t)$ is updated according to
(\ref{average_rate}). All the subchannels are allocated one-by-one
according to this rule. We will use this algorithm as benchmark in
our simulations.

\subsection{Proposed Algorithm - Delay and Rate Based Resource Allocation}

There are two main disadvantages of M-LWDF-PF algorithm. First, the
power is divided equally to over subcarriers.  Performance can be
increased by power control. Secondly, data users are much different
than video and voice in terms of QoS requirements. Therefore it is
hard to use the same metric for data and real time users. We propose
a Delay and Rate based Resource Allocation algorithm (DRA). We first
choose the users to be served in the current frame according to the
following user satisfaction value.
\begin{equation}
USV_i(t)= L_iD_i^{HOL}\log\left(1+\frac{\beta
p_i(t)h_i(t)}{N_0w_i(t)}\right)\frac{r_i^0}{R_i(t)}
\end{equation}Here $L_i=-\frac{\log(\delta_i)}{D_i^{max}}$ and
$r_i^0$ is the basic rate requirement for user i. Let $U_D$, $U_S$
and $U_V$ be the BE, Video and Voice users. Let $U_R=U_S\cup U_V$ be
the set of real time users. Let $U_E$ and $\overline{U_E}$ be the
set of users demanding elastic traffic  and the rest, respectively.

We use a simple formula to determine the fraction $F_R(t)$ of real
time users scheduled in each time slot,
\begin{equation}
F_R(t)=\frac{1}{|U_R|}\sum_{i\in U_R}I(q_i(t)>0.5D_i^{max}r_i^0)
\end{equation}
Here $q_i(t)$ is the queue size in bits and $0.5D_i^{max}r_i^0$
denotes a queue size threshold in bits and $I(.)$ is the indicator
function taking value one if the argument inside is true. As more
users exceed this threshold, more fraction of real time users are
scheduled. For data users, the BS simply chooses a fraction of 0.2
of users.

\section{Joint Power and Bandwidth Allocation}

After the users are chosen, joint power and bandwidth allocation is
performed. Let $U_D'$, $U_S'$ and $U_V'$ be the chosen users that
belong to all three traffic classes. The algorithm is as follows:

\subsection{Basic Rate Allocation for Real Time Users} For the
selected real time users ($i\in U_R'$) the rate requirements are
determined first. Rate requirement for real time user $i$ is,
\begin{equation}
r_i^c(q_i(t),\omega_i(t))=\left(\frac{q_i(t)}{T_s},\frac{r_i^0}{\omega_i(t)},
\right), ~i\in U_R'\label{Ch3:required_rate}
\end{equation}Here $q_i(t)$ is the queue size and $\omega_i(t)$ is the transmission frequency of user
i, which is updated as follows:
\begin{equation}\omega_i(t)=\alpha_i
\omega_i(t-1)+(1-\alpha_i)I(r_i(t)>0),\end{equation}where
$I(r_i(t)>0)$ is the function that takes value one if the node
receives packets in time slot $t$, zero otherwise.  Therefore this
frequency decreases if the node transmits less and less frequently.
Using this frequency expression in the basic rate function, we
compensate for the lack of transmission in the previous time slots
possibly due to bad channel conditions.

For the chosen real time users with non-elastic traffic ($i\in
\overline{U_E}\cap U_R'$) basic resource allocation is enough to
support the session. For these users we allocate the basic resource
as follows, and don't include them in the rate allocation which will
be defined later. First, the nominal SNR $\gamma_i^0$ is determined
according to the uniform power per bandwidth allocation as
$\gamma_i^0=\frac{Ph_i(t)}{N_0W}$. Then $\gamma_i^0$ is quantized by
decreasing $\frac{Ph_i(t)}{N_0W}$ to the closest SNR level in
Section \ref{Ch3:system_model}. If $\gamma_i^0$ is smaller than the
smallest SNR level, then the ceiling is taken. Based on this nominal
SINR, nominal bandwidth efficiency $S_i^0(t)$ (in bps/Hz) is
determined again using the values above. Using this basic rate and
the nominal bandwidth efficiency, basic bandwidth for non-elastic
traffic is determined as $w_i^{min} = \frac{r_i^{min}(t)}{S_i^0(t)},
i \in \overline{U_E}\cap U_R'$. Then this bandwidth is quantized to
a multiple of subchannel bandwidth by $w_i^{min}=\max(1,\lfloor
w_i^{min}\rfloor)W_{sub}$. Minimal power for this user is then
$p_i^{min}=\gamma_i^0 w_i^{min}N_0/h_i(t), ~\forall i\in
\overline{U_E}\cap U_R'$. Hence $p_i=p_i^{min}$ and $w_i=w_i^{min}$
for these users. \footnote{After the basic allocation, if the total
bandwidth or power is greater then the available resource, the user
with the largest power is chosen, bandwidth is decreased by one
subchannel and the power is also decreased in order to keep the SINR
fixed. This process is continued until the total bandwidth and power
for voice and video users becomes smaller than the available
resources.}

Let the residual power and bandwidth after non-elastic real time
traffic allocations be $P'=\sum_{i\in \overline{U_E}\cap
U_R'}p_i^{min}$ and $W'=\sum_{i\in \overline{U_E}\cap
U_R'}w_i^{min}$. For real time users with elastic traffic ($i\in
U_R'\cap U_E$) we include the basic rate as a constraint in joint
residual bandwidth-power allocation, which will be explained next.

\subsection{Proportional Fair Resource Allocation for Data and
Video Streaming} At this stage the residual power ($P'$) and
bandwidth ($W'$) is allocated among the chosen  users demanding
elastic traffic in a proportional fair manner. The PF resource
allocation problem in (\ref{Ch3:PF_allocation}) is solved among the
chosen streaming and data users.

Find ($\mathbf{p^*,w^*}$) such that:
\begin{equation}
\max_{\mathbf{p},\mathbf{w}} \prod_{i\in
  U_E\cap(U_R'\cup U_D')}{\left(w_i\log\left({1+\frac{p_i}{n_iw_i}}\right)\right)^{\phi_i}}
\label{Ch3:PF_allocation}\end{equation}subject to,
\begin{eqnarray}
w_i\log\left({1+\frac{p_i}{n_iw_i}}\right)&\geq&r_i^{min},~\forall i\in U_E\cap U_R'\label{Ch3:rate_const}\\
\sum_{i\in U_E\cap(U_R'\cup U_D')}{p_i}&\leq& P' \label{Ch3:power_const}\\
\sum_{i\in U_E\cap(U_R'\cup U_D')}{w_i}&\leq& W' \label{Ch3:band_const}\\
p_i, w_i &\geq& 0, \forall i \in U_E\cap(U_R'\cup
U_D')\label{Ch3:gr_than_zero}
\end{eqnarray}

Here log-sum is written as a product.  The above problem is a convex
optimization problem with a concave objective function and convex
set \cite{Boy04}. In this optimization we also included the
parameter $\phi_i$, which depends on the traffic type. Since data
users typically can tolerate more rate and video users are already
allocated basic bandwidth, we can give higher $\phi_i$ for data
users.  We can solve this problem using the Lagrange multipliers.

\subsection{Bandwidth and SINR quantization and Reshuffling}

After the resources are allocated, first the bandwidth for data and
video streaming users is quantized as $ w_i=\max(1, \lfloor
w_i\rfloor) W_{sub} $. Then the SINR is quantized and transmit power
is determined. Unlike best effort transmission, queue size plays an
important role in real time transmissions. As a result of the above
optimization some streaming time users may get more rates than that
is enough to transmit all bits in the queue. Some of the bandwidth
is taken from video users in order to obey this queue constraint.
After these modifications, if the total bandwidth is greater than
the available, then the user with the highest power is found and its
bandwidth decreased. Power is recalculated in order to keep the SINR
fixed. This process is continued until bandwidth constraint is
satisfied. If total power is still greater than the available then
again choosing the user with highest power and decreasing bandwidth,
power constraint is satisfied. If after these processes there is a
leftover bandwidth, then choosing the user that has the highest
channel a subchannel is added and power is increased accordingly (if
there is enough power to do so). If there is some leftover power,
then starting from the user with lower channel gains, SINR is
boosted to the next power level (if there is enough power to do so).
For the real time users we don't increase bandwidth or power if
there isn't enough buffer content.
\section{Numerical Evaluation}
For the numerical evaluations we divide the users to 5 classes
according to the distances, 0.3,0.6,0.9,1.2,1.5 km. There are equal
number of users at each class. We use the parameters in Table
\ref{simulation_parameters}.
\begin{table}[htb]
\footnotesize \centering
\begin{tabular}{|c|c|}
\hline
 \textbf{Parameter}&\textbf{Value}\\\hline\hline
Cell radius & 1.5km\\\hline

User Distances & 0.3,0.6,0.9,1.2,1.5 km \\\hline

Total  power (P)&20 W \\\hline

Total bandwidth (W) &10 MHz\\\hline

Frame Length & 1 msec \\\hline

Voice Traffic & CBR 32kbps \\\hline

Video Traffic &  802.16 - 128kbps \\\hline

Best effort File &  5 MB \\\hline

AWGN p.s.d.($N_0$)&-169dBm/Hz  \\\hline 

Pathloss exponent ($\gamma$)& 3.5\\\hline

$\psi_{DB}\sim N(\mu_{\psi_{dB}},\sigma_{\psi_{dB}})$ & N(0dB,8dB)
\\\hline

 Coherent Time (Fast/Slow) & (5msec/300msec.)  \\\hline

Pathloss(dB, d in meters)& $-31.5- 35\log_{10}d+\psi_{dB}$\\\hline
\end{tabular}
\caption{Simulation Parameters}\label{simulation_parameters}
\end{table}

We performed the simulations using MATLAB. We compared our algorithm
with the benchmark M-LWDF algorithm with proportional fairness.
Delay exceeding probability is taken as $\delta_i=0.05$ for all
users. The traffic and resource allocation parameters are listed in
Table \ref{parameters}.  Since we choose data users separately from
others, the parameters $L_i$ and head of line delay $D_i^{HOL}$ are
not used for data users.

\begin{table}[htb]
\footnotesize \centering
\begin{tabular}{|c|c|c|c|c|c|c|c|}
\hline
 \textbf{Traffic}&$r^{0}(kbps)$&$r^{max}(kbps)$&$D^{max}(s)$&$L_i$ & $\phi_i$ & $\alpha_i$\\\hline\hline
VoIP & 32& 32 &0.1& 13& -& 0.98\\\hline

Streaming&128& 1024 &0.4& 3.25&1 & 0.995\\\hline


BE &0& $\infty$ & 2 & 0.65&-&0.998\\\hline \end{tabular}
\caption{Minimum required and maximum sustained rates for different
types of traffic.}\label{parameters}
\end{table}

The measured performance metrics are $95^{th}$ percentile delay for
real time users and total throughput for data users. We will observe
these parameters with respect to number of video users. For the
delay, we observe the users in the range 0.3-1.2 separately as
\textit{good} users and the ones at 1.5km as \textit{bad} users.

\subsection{Fixed Rate Video Traffic}
In the first part of the simulations we considered the video traffic
rate fixed at 128kbps and treated it as non-elastic.  We consider
CBR voice traffic, where a fixed length packet arrives periodically.
For the Video traffic we used the model in IEEE 802.16e system
evaluation methodology. Packet lengths, and interarrival times
truncated Pareto distributed such that average rate is 128kbps. For
the BE traffic we assume that there are unlimited number of packets
in the queue.





\begin{figure}[htb]
\begin{center}
\epsfig{file=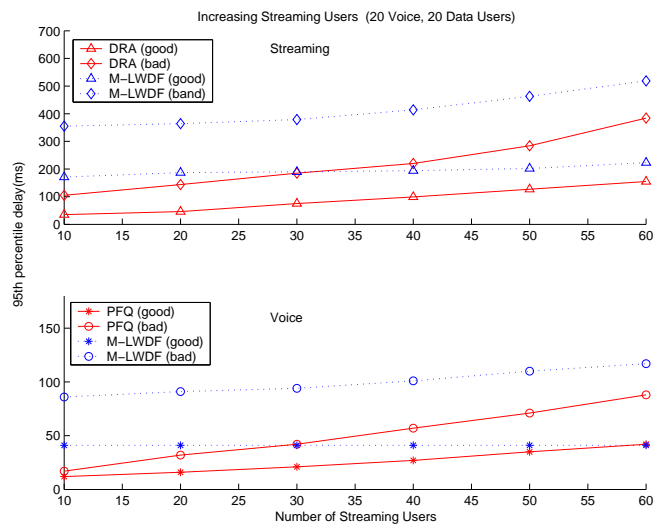, height=2.75in}
\end{center}
\caption{95 percentile delay(msec) vs. number of video
users}\label{95_percent_video}
\end{figure}

In Figure \ref{95_percent_video}, we plotted the 95 percentile
delays of real time users vs increasing number of video users. For
this simulation we kept the number of data and Voice users fixed at
20. Again we observe that $95^{th}$ percentile delay for video users
increases exponentially with number video users, while delays for
the users at the edge is within the acceptable range for DRA unlike
M-LWDF.

\begin{figure}[htb]
\begin{center}
\epsfig{file=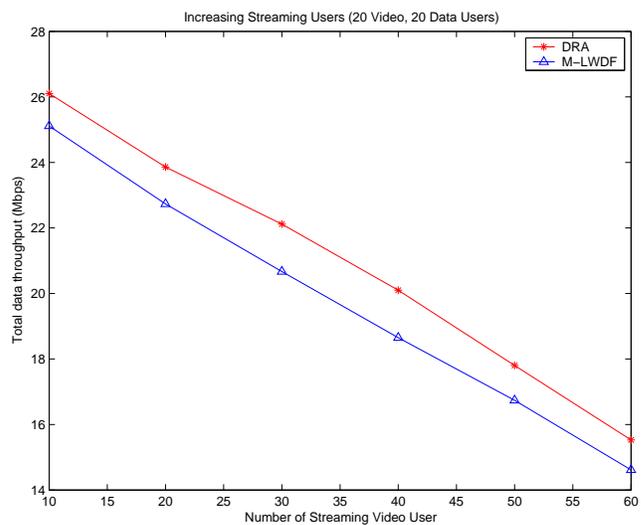, height=2.75in}
\end{center}
\caption{Total throughput(Mbos) vs. number of video
users}\label{throughput_video}
\end{figure}

In figure \ref{throughput_video} we see that total data rate
decreases linearly with increasing video users. Data performance of
DRA is again better than M-LWDF.

\begin{figure}[htb]
\begin{center}
\epsfig{file=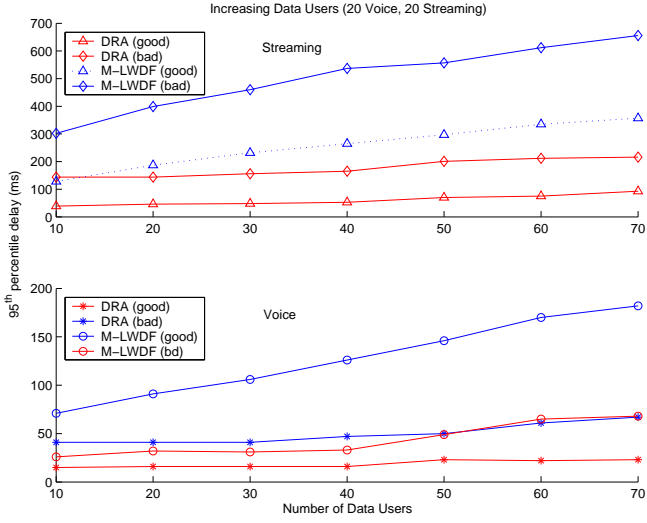, height=2.75in}
\end{center}
\caption{95 percentile delay(msec) vs. number of data
users}\label{95_percent_ftp}
\end{figure}

In Figure \ref{95_percent_ftp}, $95^{th}$ percentile delay for video
and voice users are plotted for increasing number of data users. The
number of Streaming and Voice users are kept fixed at 20. We observe
a linear increase in the delay w.r.t. number of data users with
M-LWDF.  The delay increase is negligible for DRA.

\subsection{Elastic Video Traffic}
In the second part of the simulations we considered video traffic
rate that varies with packet delays.   We implemented a simple rate
control scheme that looks at the average head of line packet delay
and increases or decreases input rate according to a threshold
policy. We defined rate levels $r_i^0\lambda_i$, ($\lambda_i\in
\{1,2,\ldots,8\}$) that are integer multiples of 128kbps.
Interarrival times are the same for level $1$ and $k$, however for
level $k$ packet size is $k$ times larger for each packet. For each
user $i\in U_E\cap U_R$ and at each update instant.
\begin{itemize}
\item if $\overline{D_i^{HOL}}(t)<0.125D_i^{max}$ then $ \lambda_i = \min\{\lambda_i+1,\lambda^{max}\}$
\item if $\overline{D_i^{HOL}}(t)>0.25D_i^{max}$ then $ \lambda_i = \max\{\lambda_i-1,1\}$
\item else, $\lambda_i = \lambda_i$
\end{itemize}Here $\overline{D_i^{HOL}}(t)$ denotes mean HOL packet
delay in the last $400$ frames. The updates are made at each $200$
frames.

\begin{figure}[htb]
\begin{center}
\epsfig{file=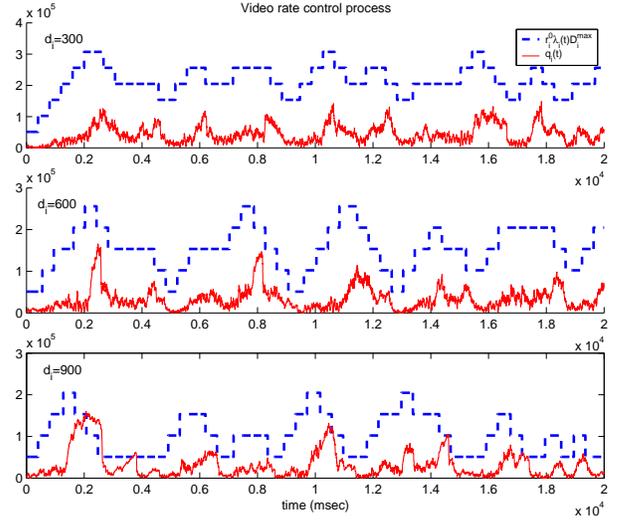, height=2.75in}
\end{center}
\caption{Evolution of Video rate along with queue sizes for users at
300, 600 and 900meters}\label{Ch3:rate_evolution}
\end{figure}

\begin{figure}[htb]
\begin{center}
\epsfig{file=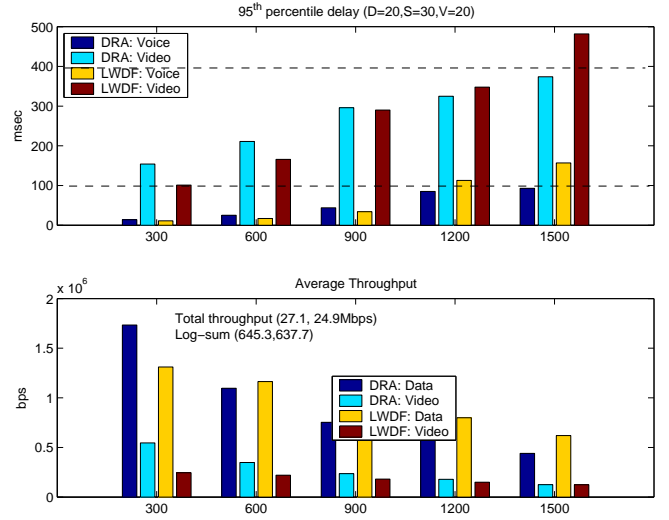, height=2.75in}
\end{center}
\caption{$95^{th}$ percentile delay and average throughput for users
at different distances.}\label{Ch3:delay_throughput}
\end{figure}

Figure \ref{Ch3:rate_evolution} shows the evolution of rate levels
along with queue sizes for video users at distances 300, 900 and
1500 meters. We observe that users closer to the BS can achieve
higher rates.
In Figure \ref{Ch3:delay_throughput} we observe the comparison of
delay and throughput for the DRA and LWDF schemes.We see that DRA
system satisfies delay constraints for voice users unlike LWDF.  As
for throughput, we see that DRA can provide significantly better
throughput for video users at all distances.  Total data/video
throughput and log-sum throughput (proportional fairness) is also
better for DRA scheme.

\bibliographystyle{unsrt}
\bibliography{adaptive_metric_conference}
\end{document}